\documentclass[preprint2,epsf,epsfig,graphics]{emulateapj}
\usepackage[breaklinks,colorlinks, urlcolor=blue,citecolor=blue,linkcolor=blue]{hyperref}
\usepackage{graphicx}
\usepackage{color}
\usepackage{mathtools}
\usepackage{amsmath}
\usepackage{color}
\def\dterm{$D$-term}

\shorttitle{Polarized foregrounds for EoR experiments \sc{i}}
\shortauthors{Kohn et al.}

\begin{document}
\title{Constraining polarized foregrounds for EoR experiments \sc{i}: 2D power spectra from the PAPER-32 imaging array}

\author{S.~A. Kohn\altaffilmark{1}}
\author{J.~E. Aguirre\altaffilmark{1}}
\author{C.~D. Nunhokee\altaffilmark{2}}
\author{G. Bernardi\altaffilmark{3,2,4}}
\author{J.~C. Pober\altaffilmark{5}}
\author{Z.~S. Ali\altaffilmark{6}}
\author{R.~F. Bradley\altaffilmark{8,9,10}}
\author{C.~L. Carilli\altaffilmark{11,12}}
\author{D.~R. DeBoer\altaffilmark{6}}
\author{N.~E. Gugliucci\altaffilmark{13}}
\author{D.~C. Jacobs\altaffilmark{14}}
\author{P. Klima\altaffilmark{9}}
\author{D.~H.~E. MacMahon\altaffilmark{7}}
\author{J.~R. Manley\altaffilmark{3}}
\author{D.~F. Moore\altaffilmark{1}}
\author{A.~R. Parsons\altaffilmark{6, 7}}
\author{I.~I. Stefan\altaffilmark{12}}
\author{W.~P. Walbrugh\altaffilmark{3}}

\email{saulkohn@sas.upenn.edu}
\altaffiltext{1}{Department of Physics and Astronomy, University of Pennsylvania, Philadelphia, PA, USA}
\altaffiltext{2}{Department of Physics and Electronics, Rhodes University, Grahamstown, South Africa}
\altaffiltext{3}{SKA South Africa, Pinelands, South Africa}
\altaffiltext{4}{Harvard-Smithsonian Center for Astrophysics, Cambridge, MA, USA}
\altaffiltext{5}{Department of Physics, Brown University, Providence, RI, USA}
\altaffiltext{6}{Astronomy Department, University of California, Berkeley, CA, USA}
\altaffiltext{7}{Radio Astronomy Laboratory, University of California, Berkeley, CA, USA}
\altaffiltext{8}{Department of Electrical and Computer Engineering, University of Virginia, Charlottesville, VA, USA}
\altaffiltext{9}{National Radio Astronomy Observatory, Charlottesville, VA, USA}
\altaffiltext{10}{Department of Astronomy, University of Virginia, Charlottesville, VA, USA}
\altaffiltext{11}{National Radio Astronomy Observatory, Socorro, NM, USA}
\altaffiltext{12}{Cavendish Laboratory, Cambridge, UK}
\altaffiltext{13}{Saint Anselm College, Manchester, NH, USA}
\altaffiltext{14}{School of Earth and Space Exploration, Arizona State University, Tempe, AZ, USA}

\begin{abstract}
Current-generation low frequency interferometers constructed with the objective of 
detecting the high-redshift 21\,cm background, aim to generate power spectra of the brightness-temperature contrast of neutral hydrogen in primordial intergalactic medium. 
Two-dimensional power spectra (power in Fourier modes parallel and perpendicular to the line of sight) formed from interferometric visibilities have been shown to delineate a boundary between spectrally-smooth foregrounds (known as the \textit{wedge}) and spectrally-structured 21\,cm background emission (the \textit{EoR-window}).
However, polarized foregrounds 
are known to possess
spectral structure due to Faraday rotation, which can 
leak into the EoR window. 
In this work, we create and analyze 2D power spectra from the PAPER-32 imaging array in Stokes I, Q, U and V. 
These allow us to observe and diagnose systematic effects in our calibration at high signal-to-noise within the Fourier space most relevant to EoR experiments.
We observe well-defined windows in the Stokes visibilities, with Stokes Q, U and V power spectra sharing a similar wedge shape to that seen in Stokes I.  With modest polarization calibration, we see no evidence that polarization calibration errors move power outside the wedge in any Stokes visibility, to the noise levels attained.  Deeper integrations will be required to confirm that this behavior persists to the depth required for EoR detection.

\end{abstract}

\setcounter{footnote}{0}

\section{Introduction}
\label{sec:intro}
The Epoch of Reionization (EoR) marks an era in which the intergalactic medium (IGM) underwent a phase 
transition from neutral to ionized. Between redshifts $\sim8<z<200$, the first stars and black holes formed from clouds of hydrogen left over from recombination, and could, for the first time since the last-scattering surface, ionize neutral hydrogen. Due to the huge range in redshift (i.e. volume) that this process occurs over, the ability to intensity-map neutral hydrogen via its 21\,cm ($\approx$\,1420\,MHz) hyperfine transition over cosmic time (i.e. different redshifted frequencies) as been recognized as one of the most powerful probes of cosmic dawn \citep[e.g.][]{Furlanetto.06}. 

Many experiments seek to constrain the power spectra of 21\,cm brightness across different redshifts using low-frequency interferometers. These instruments include the GMRT \citep{Paciga.11}, LOFAR\footnote{\url{www.lofar.org}} \citep{vanHaarlem.13}, the MWA\footnote{\url{www.mwatelescope.org}} \citep{Tingay.13} and the Precision Array for Probing the Epoch of Reionization; PAPER\footnote{\url{eor.berkeley.edu}} \citep{Parsons.10}. To date, PAPER has imposed the strongest limits on the 21\,cm power spectrum at $z = 7.7, 8.4$~and~$10.3$, respectively \citep[]{Parsons.14,Ali.15,Jacobs.14,Moore.15}.

The biggest challenge that all of these experiments share is the overwhelming power of foreground emission, orders of magnitude brighter than the reionization signal \citep[e.g.][]{Bernardi.09,Pober.13,Dillon.14}. Foreground emission, however, is expected to be intrinsically spectrally smooth over tens of MHz, whereas the inherent cloudiness of neutral hydrogen before and during reionization is expected to vary over a few MHz scale. This different spectral behavior still constitutes the 
most powerful leverage to separate the two signals.
Looking at their footprint in power spectrum space, foregrounds are expected to be confined into a `wedge'-shaped region, leaving an open \textit{EoR window} for interferometric reionization studies to take advantage of \citep[e.g.][]{Datta.10, Morales.12, Trott.12, Pober.13, Pober.14, Liu.14a, Liu.14b, Dillon.15b, Dillon.15a, Nithya.15b, Nithya.15a}. 
PAPER EoR observations implement foreground avoidance strategies, filtering-away the wedge. However, by combining knowledge of the wedge with foreground removal  (modelling and subtracting sources from the data), significantly more EoR signal could be recovered, especially if foreground spectra deviate from perfect smoothness \citep{Chapman.14}.

Polarized synchrotron emission from our own Galaxy and from extragalactic radio sources is not expected to be confined within the `wedge'-shaped region due to spectral structure from Faraday rotation, and imperfect calibration may leak 
such emission into the total intensity, representing a potentially serious contamination to the EoR signal \citep{Bernardi.10,Jelic.10,Moore.13}. Recent observations in the $100-200$MHz~band
have revealed that Galactic diffuse polarized synchrotron radiation seems to be ubiquitous \citep{Bernardi.09,Bernardi.10,Jelic.14}, with peak emission up to 15~K \citep{Jelic.15}, but find a dearth of polarized point sources.  A study of the depolarization of point sources by \citet{Farnes.14} showed a systematic trend for depolarization of steep-spectrum point sources as frequency decreased, resulting in very low polarization fractions ($<<1\%$) below 300 MHz.
\cite{Bernardi.13} surveyed the largest sky area in polarization to date, showing that polarized extragalactic sources are significantly depolarized with respect to higher frequencies, detecting only one source with polarization fraction greater than 2\% over $\sim 2400^\circ$. 
This low level of intrinsic polarization appears to be corroborated by recent results from LOFAR \citep{Asad.15}.
\cite{Bernardi.13} further showed that diffuse polarized emission largely occurs at rotation measure (RM) values~$< |10-15|$~rad~m$^{-2}$ that are expected to contaminate $k$ modes up to $k \sim 0.05$~Mpc$^{-1}$ at $z = 8$ \citep{Moore.13}.

As PAPER and other instruments are beginning to reach levels of sensitivity comparable to the expected reionization signal \citep[e.g.][]{Lidz.08}, polarized leakage may become a dominant systematic. 
For PAPER, the effective polarization leakage over its long season-averages has been limited to be below the current upper limits on the 21 cm power spectrum at the relevant $k$-modes 
\citep{Moore.15}. 
\citet{Asad.15} found evidence of very small amounts of polarized leakage (0.2--0.3\%) into an EoR window constructed from the central 4$^{\circ}$ of the LOFAR 3C196 field.

As the PAPER collaboration looks forward to analyzing data from the instrument's final configuration with 128 dual-polarization dipole antennae in a highly redundant array configuration, and as the Hydrogen Epoch of Reionization Array (HERA)\footnote{\url{www.reionization.org}} is now under construction \citep[see, e.g., ][]{Ewall-Wice.16, Neben.16}, this level of leakage 
may become perceivable in the power spectra measured by these instruments.

It is therefore timely to constrain both instrumental and astrophysical polarization to allow instruments such as PAPER and HERA to make the best possible measurements of the 21\,cm power spectrum. In this paper, we aim to constrain the polarized foreground footprint in the two dimensional power spectrum and its correlation with the total intensity wedge shape, and quantify the effects of calibration within the wedge for all four Stokes parameters.

This work is the first in a series of papers that will study the impact polarization leakage for the PAPER and HERA experiments. This paper details how 2D polarized power spectra created from visibilities using the delay transform, with basic direction-independent calibration, can be used to diagnose instrument systematics. 

The layout of this paper is as follows.  In Section~\ref{sec:obs} we provide a brief description of the PAPER array in its imaging configuration, the data from which this paper is based, and describe its calibration and reduction. We also describe the method used to create 2D power spectra in this section. We analyze the power spectra in Section~\ref{sec:res}, and discuss the implications of our findings and conclude in Section~\ref{sec:disc}.

\section{Observations}
\label{sec:obs}

The PAPER 32-antenna array relied on its highly redundant configuration in order to take the measurements resulting in the strong upper limits on the 21\,cm power spectrum \citep{Parsons.14, Jacobs.14, Moore.15}. However, for three nights in September 2011 the 32 elements were reconfigured into an polarized imaging configuration. We present measurements taken overnight on September 14--15, 2011 over  local sidereal times (LSTs) 0h--5h.

\subsection{Observation and reduction}

\begin{figure}
\centering
\includegraphics[width=0.9\columnwidth]{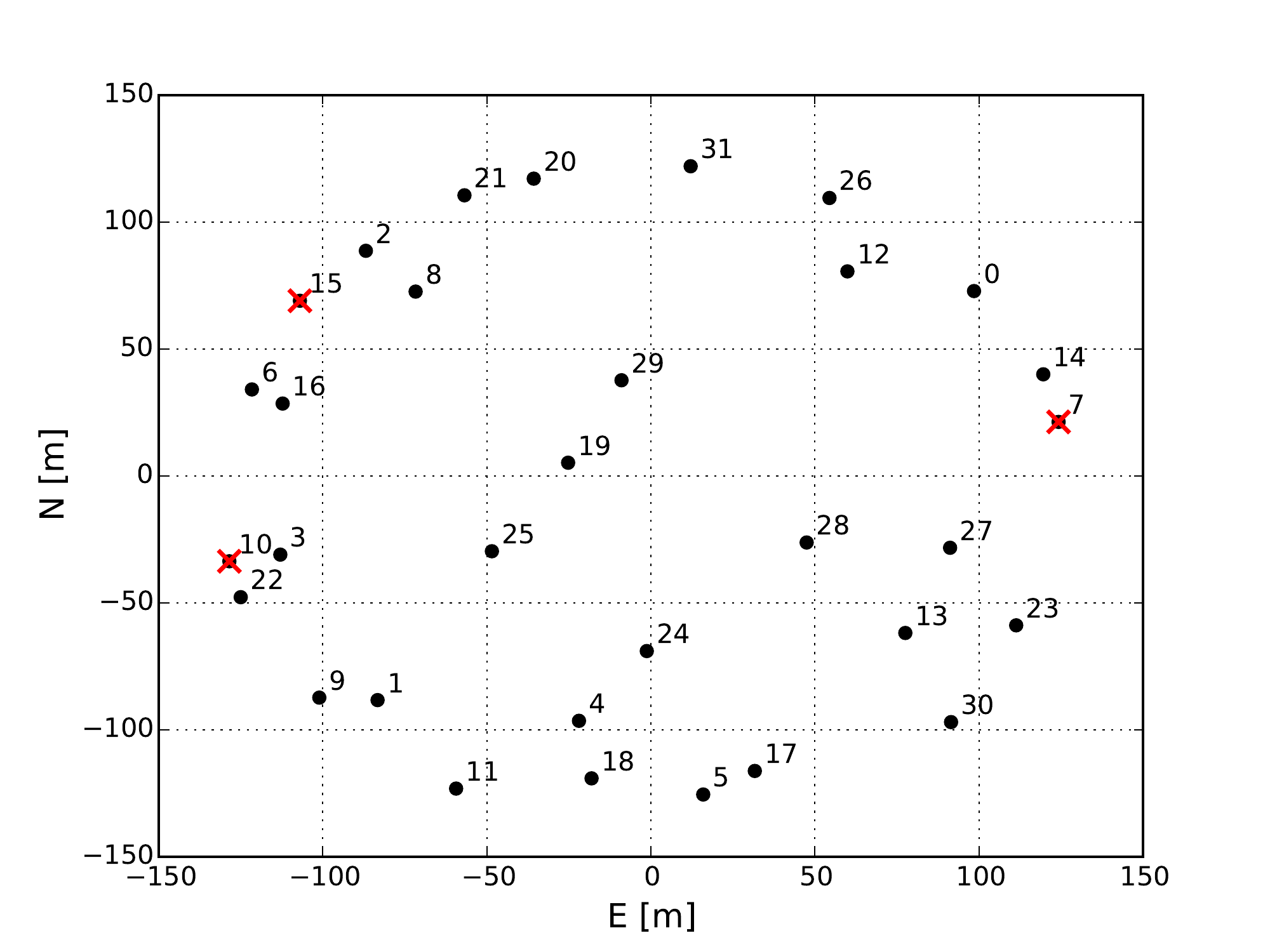}
\includegraphics[width=0.9\columnwidth]{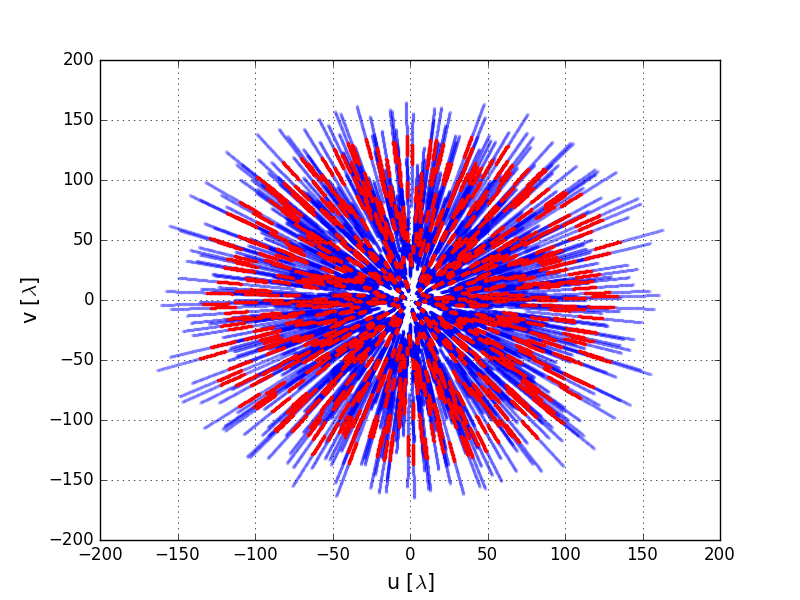}
\caption{The PAPER-32, dual-pol antenna imaging configuration (top). They were arranged in a pseudo-random scatter within in a $\sim$300\,m diameter circle to maximize instantaneous $uv$ coverage (bottom). $uv$ coverage is shown for 100--200\,MHz over 203 channels in blue, and 146--166\,MHz over 20 channels in red (the latter being the frequencies used in our power spectrum analysis). Malfunctioning antennae identified during calibration are overlaid with red crosses (and are excluded from the $uv$ coverage map).\\}
\label{fig:config}
\end{figure}

Antennae were arranged in a pseudo-random scatter within a 300\,m-diameter circle, the layout of which is shown in Figure~\ref{fig:config}. This allowed us to obtain resolutions between 15' and 25' across the bandwidth (100--200 MHz nominally, although in reality this extends 110--185 MHz due to band edge effects and VHF TV). Drift-scan visibilities were measured every 10.7 s, and divided into datasets about 10 minutes in length. We express an interferometric visibility $V^{pq}_{ij}$ between antennae $i$ (with dipole arm $p$, which can be $x$ (East-West) or $y$ (North-South) for PAPER dipoles), and $j$ (with dipole arm $q$), in directional cosines $l$ and $m$ for frequency $\nu$ at time $t$, as:

\begin{equation}
\begin{aligned}
V_{ij}^{pq}(\nu,t) = g^p_i g^{q*}_j \exp(-2\pi i \nu \tau_{pq}) \times \\ \int d\Omega \, A^{pq}(\Omega, \nu)
S(\Omega, \nu) \exp\left(\frac{ -i\nu}{c} \vec{b}(t) \cdot \hat{s}(\Omega)\right)
\end{aligned}
\label{eq:visibility}
\end{equation}

where the $g$ terms represent the complex gains for each antenna and dipole arm, $A^{pq}$ is the polarized beam and $S$ is the sky. The product $\vec{b}(t) \cdot \hat{s}(\Omega)$ represents the projection of the baseline between $i$ and $j$ with respect to an arbitrary location on the sky. 
The motivation for including the term for the delay between dipole arms $p$ and $q$, $\tau_{pq}$, is given in Section \ref{subsubsec:polcal}.  This delay is clearly zero if $p=q$.

Visibilities were obtained from correlating both $x$ and $y$ dipoles, forming V$^{xx}$, V$^{xy}$, V$^{yx}$ and V$^{yy}$. Frequencies from 100 to 200\,MHz were sampled into 2048 channels.
Data were delay-filtered to 203 frequency channels \citep[see the Appendix of][]{Parsons.14}.  Cross-talk was modeled and removed by subtracting the average power over the 5 hours of observation, which extended across LST=0h--5h. An initial RFI-flagging removed any outliers more than $6\sigma$ from a spectrally smooth profile.

\subsection{Calibration}

Calibration took place in three stages, detailed below: a first-order delay-space calibration for the initial gains and phases with respect to Pictor A, an absolute calibration using imaging with respect to Pictor A and Fornax A, and a polarimetric correction for the $\tau_{xy}$ phase term in the V$^{xy}$ and V$^{yx}$ visibilities. 
Traditional polarimetric calibration proceeds by observing a source with a known polarization angle, and solving for up to seven direction-independent terms in the Jones matrix \citep[e.g.][]{TMS, HBS-1}, as well correcting for the effects of the primary beam. 
As discussed in Section~\ref{sec:intro}, given the dearth of suitable calibrators at our observing frequencies, especially at the relatively low resolution and sensitivity of the array, we proceeded with polarized calibration using different techniques, as described in Section~\ref{subsubsec:polcal} below.

\subsubsection{Initial calibration}
A first-order gain and phase calibration was performed by a similar approach to \citet{Jacobs.13}. Each 10~minute drift-scan dataset was phased to the known position of Pictor A using {\sc aipy}\footnote{\url{https://github.com/AaronParsons/aipy}} routines.

The gain term in Equation \ref{eq:visibility} was approximated as 
\begin{equation}
 g^p_i = G^p_i \exp(-2\pi i \nu \tau_{ip})
\end{equation}
and the required delay $\tau_{ip}$ offset of the uncalibrated delay tracks to the real position on the sky solved for to obtain a phase calibration; the absolute flux calibration $G^p_i$ was found by isolating the tracks of Pictor A in delay space, and applying the required flux scale across the band \citep[for a discussion of delay-space calibration, see][and Figure~\ref{fig:delay_spectra}]{Parsons.12a}.

\subsubsection{Absolute Calibration}
\label{subsubsec:abscal}

Visibilities were converted to CASA\footnote{\url{http://casa.nrao.edu}} Measurement Sets to be further calibrated using a custom pipeline developed around CASA libraries. Snapshot images were generated for each 10~minute observation by Fourier transforming the visibilities. 
We used uniform weights and the multi-frequency synthesis algorithm to further improve the $uv$~coverage. 
Dirty images were deconvolved down to a 5~Jy threshold using the Cotton-Schwab algorithm. 
The sky model generated by the CLEAN components was used to self-calibrate each snapshot over the full bandwidth, using a frequency-independent sky-model and averaging over the 10~minute observation.
We corrected for residual cable length errors by computing antenna-based phase solutions for each frequency channel for each snapshot observation. 
After self-calibration, snapshot visibilities were again Fourier transformed into images and deconvolved down to a 2~Jy threshold to form the final sky models. 
These final sky models were used to solve for a frequency independent, diagonal, complex Jones matrix \citep{HBS-1,Smirnov.11} for each antenna in order to calibrate gain variations from snapshot to snapshot. 
We make no attempt to correct sky models for polarized primary beams and, therefore, our gain solutions incorporate both the direction independent and the direction dependent responses of the two gain polarizations. 
This is a reasonable approximation for the scope of the paper, as, eventually, wide-field polarization corrections cannot be implemented directly in the per-baseline 
power spectrum estimation (see Section~\ref{sec:res}).

The average correction in magnitude through this second-order calibration was a $\pm$6\% change for $x$ gains and $\pm$7\% for $y$ gains from those derived in the initial delay-space calibration.
If the gain on an antenna deviated by more than 30\% from image-to-image during this analysis, it was discarded from future processing stages, since it was likely malfunctioning. This was true for 3 antennae (see the top panel of Figure~\ref{fig:config}). 

\begin{figure*}
\centering
\includegraphics[width=0.4\textwidth]{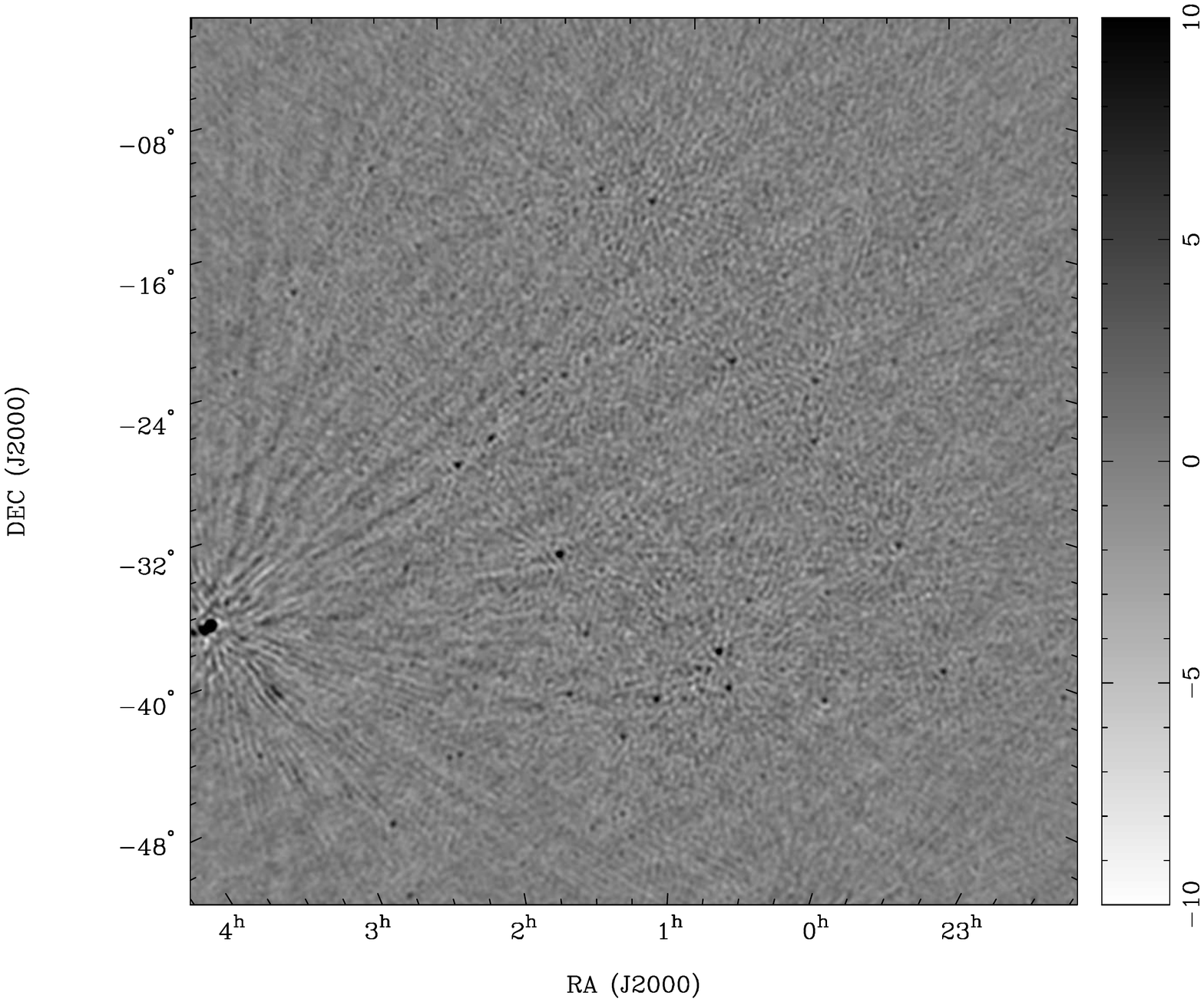}
\includegraphics[width=0.4\textwidth]{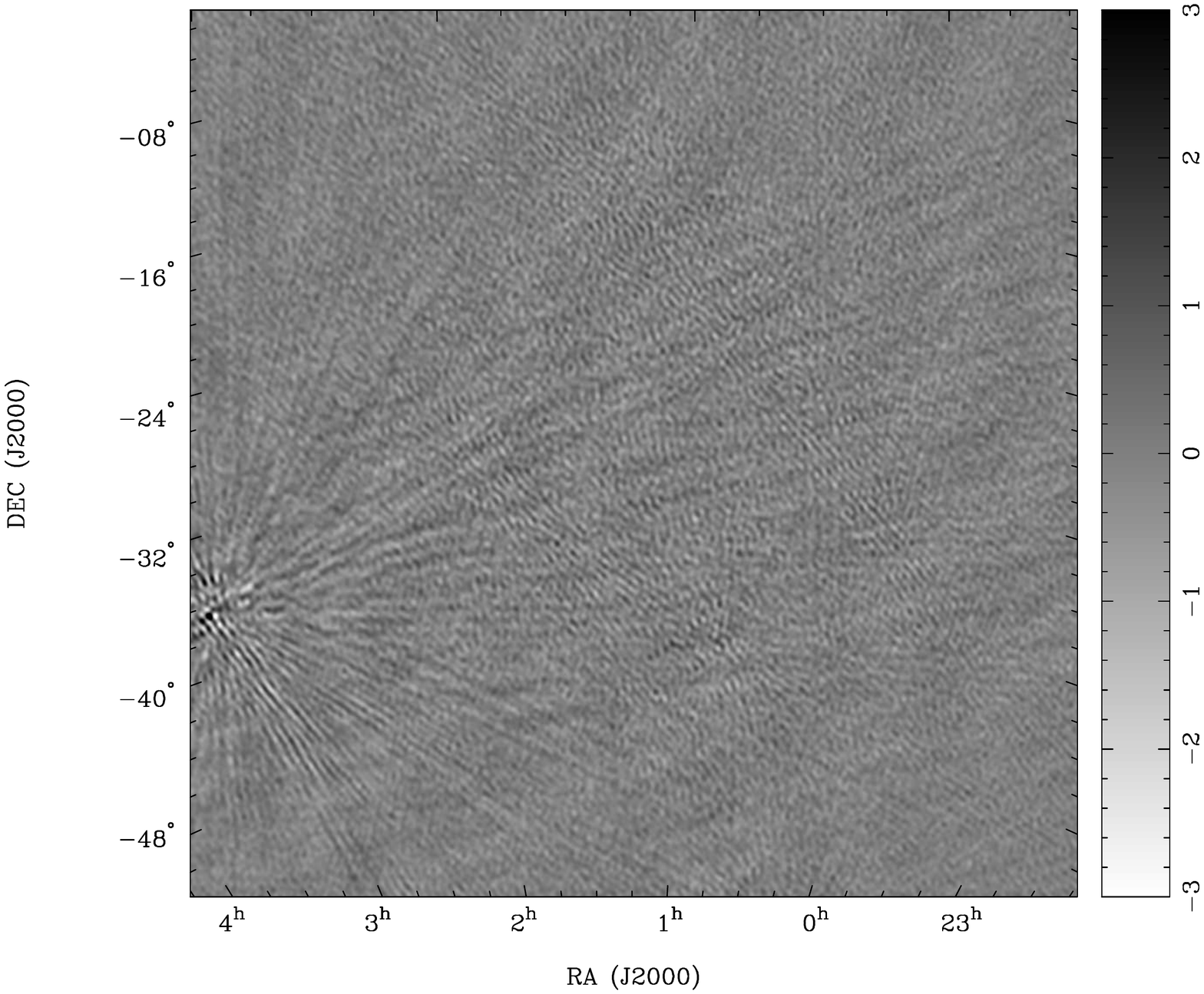}
\includegraphics[width=0.4\textwidth]{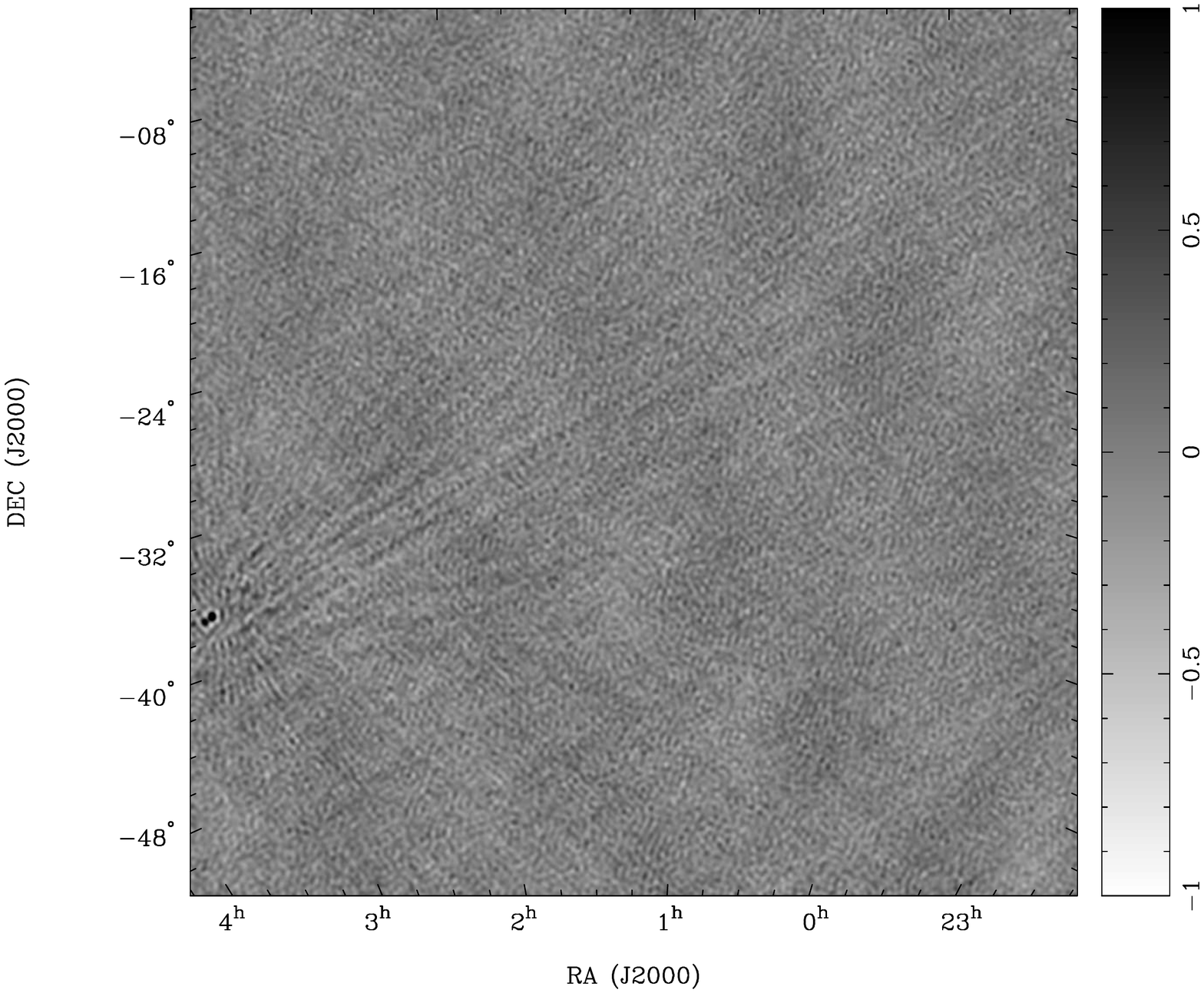}
\includegraphics[width=0.4\textwidth]{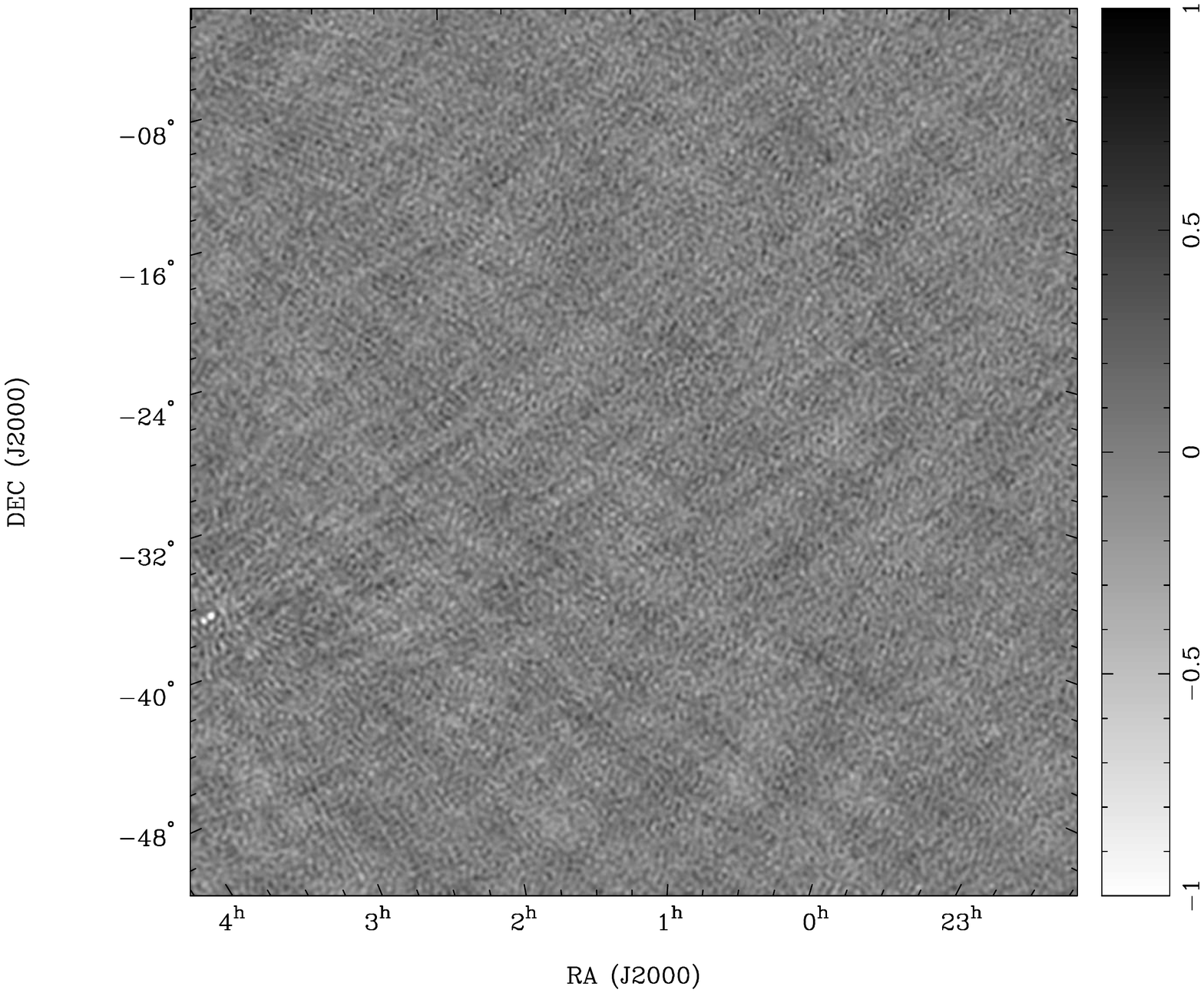}
\noindent\rule{14cm}{0.6pt}
\includegraphics[width=0.4\textwidth]{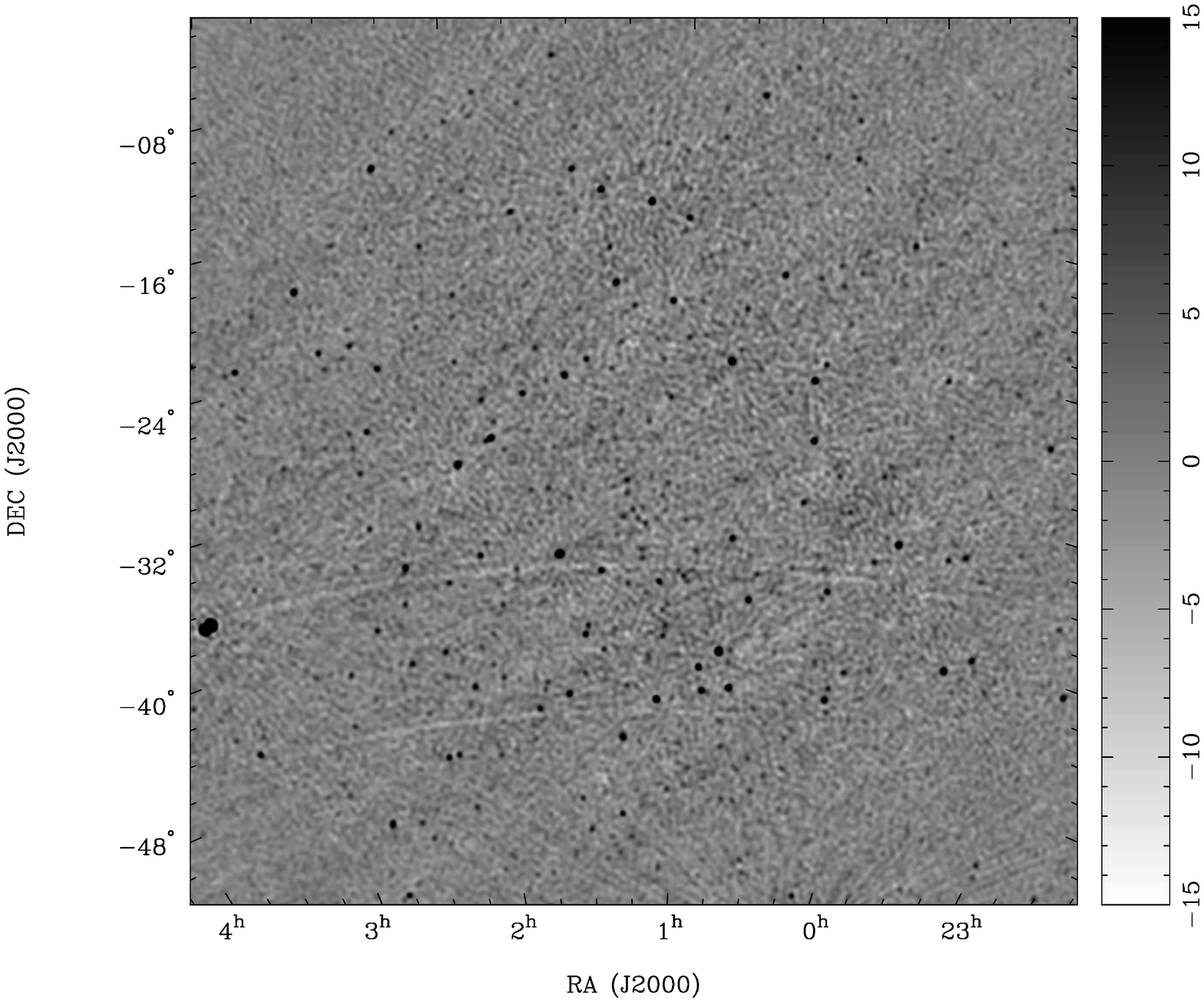}
\includegraphics[width=0.4\textwidth]{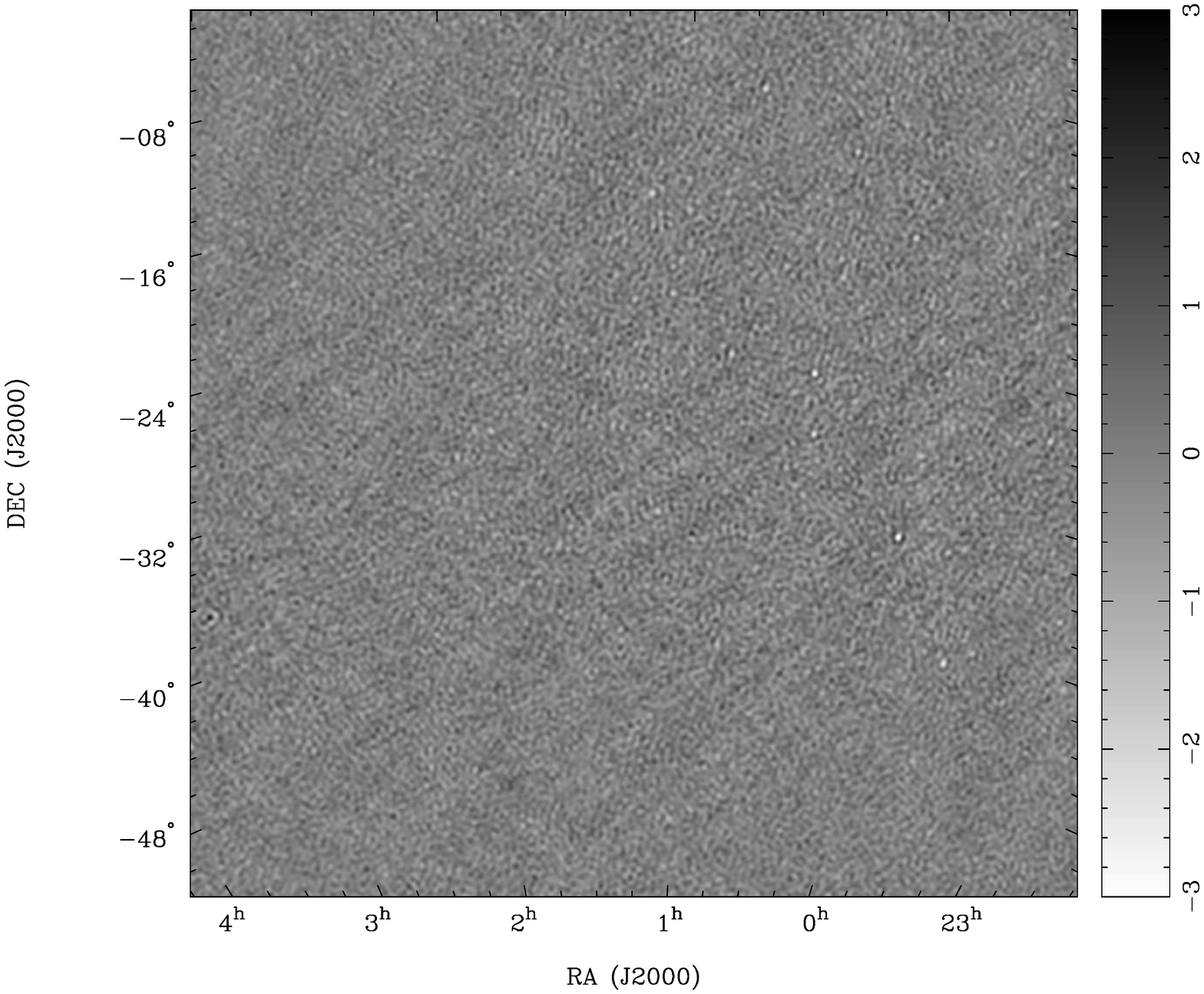}
\includegraphics[width=0.4\textwidth]{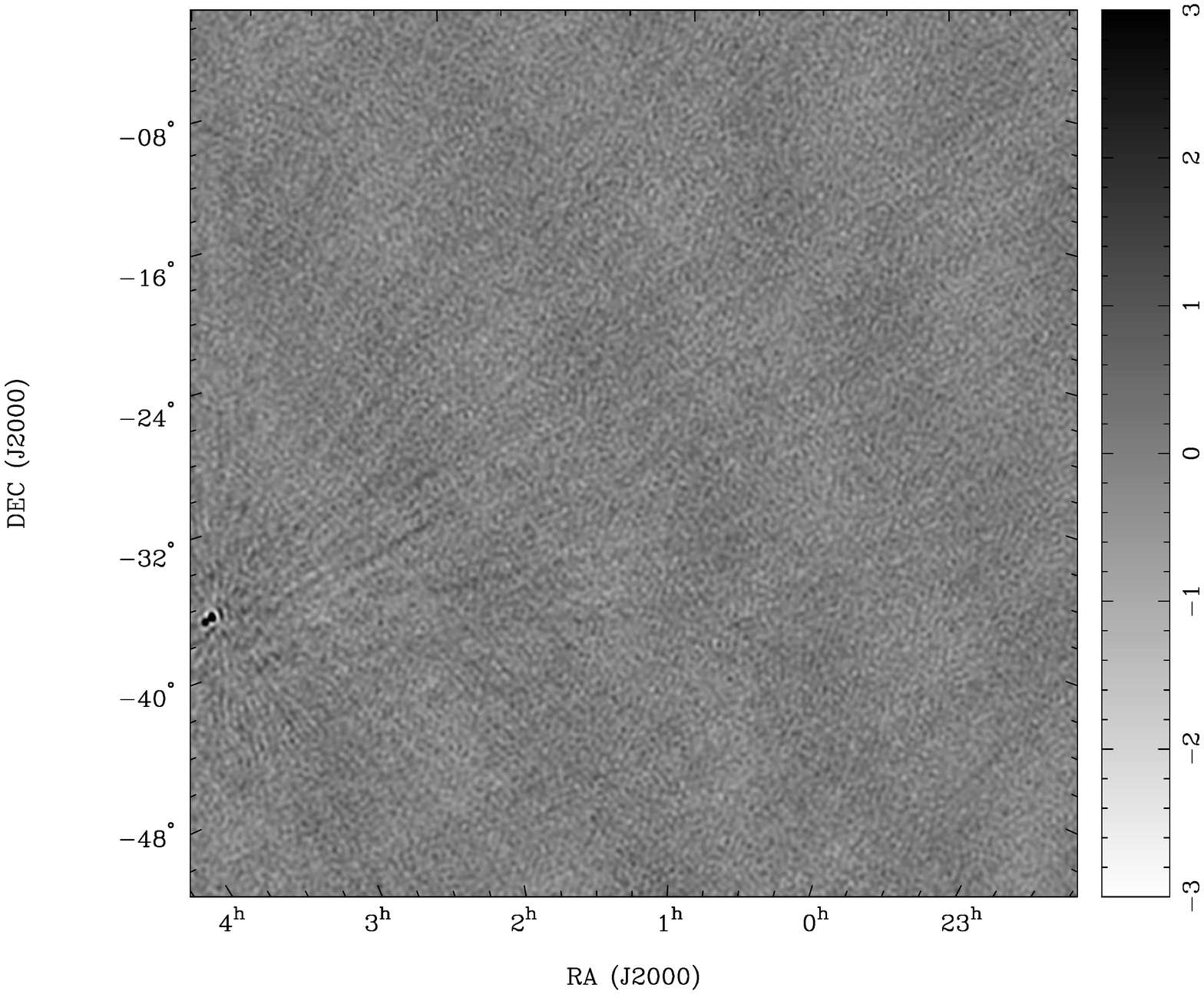}
\includegraphics[width=0.4\textwidth]{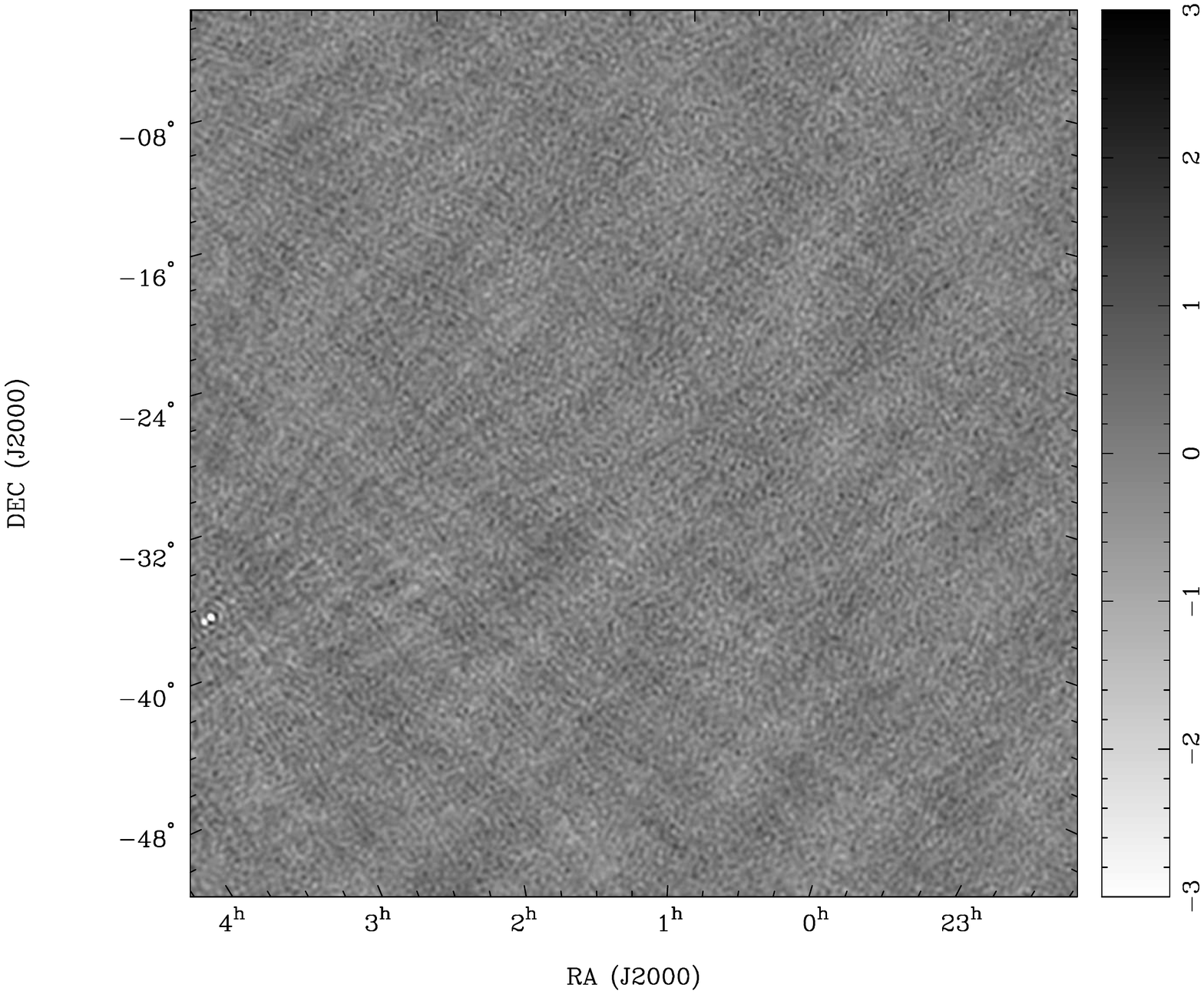}
\caption{\textit{Above:} Example of a Stokes I snapshot image (top left) with corresponding Stokes Q (top right), Stokes U (bottom left) and Stokes V (bottom right) images before absolute calibration. \textit{Below:} The same organization as above, after absolute calibration. No primary beam correction was applied. The Stokes I image was deconvolved down to 5~Jy~beam$^{-1}$ whereas the other images were not deconvolved. We note that the Stokes Q image is relatively featureless apart from a few faint sources that appear instrumentally polarized. Stokes U and Stokes V images are, instead, dominated by Fornax A that shows instrumental polarization leaked from total intensity. Units are Jy~beam$^{-1}$; note the change in scale between polarizations and calibration stages.}
\label{fig:sky_image}
\end{figure*}

The final gain amplitude calibration was carried out similarly to \citet{Ali.15}. We generated single channel images between 120 and 174 MHz for each snapshot and deconvolved each of them down to 10 Jy. For each snapshot, a source spectrum is derived for Pictor A by fitting a two dimensional Gaussian the source using the PyBDSM\footnote{\url{http://www.lofar.org/wiki/doku.php?id=public:user software:pybdsm}} source extractor \citep{pybdsm}. Spectra were optimally averaged together by weighting them with the primary beam model evaluated in the direction of Pictor A. To fit the absolute calibration, we divided the model spectrum \citep{Jacobs.13} by the measured one and fit a 6th order polynomial over the 120-174 MHz frequency range. This procedure was repeated using Fornax A with the only difference that a taper was applied to the visibilities (120\,m) in order to reduce Fornax A to a point-like source and use the model spectrum from \citet{Bernardi.13}. The best fit coefficients for Pictor A and Fornax A were averaged together to obtain the final absolute flux density calibration. Snapshots of fully CASA-calibrated data are shown in Figure~\ref{fig:sky_image}.

\subsubsection{Polarimetric factors}
\label{subsubsec:polcal}
Standard full polarization calibration involves correcting for leakage of Stokes~$I$ into the $V_{ij}^{xy}$ and $V_{ij}^{yx}$ visibilities and leakage of polarized signal into total intensity (the so called Jones $D$ matrices or \dterm s; e.g. \citet[][]{TMS, HBS-1}), and an unknown phase difference between the $x$ and $y$ feeds \citep[e.g.][]{Sault.96}. 

We attempt no $D$ matrix calibration in this paper, as there is not a dominant source to be used for such calibration: the limited sensitivity of our observations does not offer good signal-to-noise ratio on PMN~J0351-2744, the only  polarized source at low frequencies known so far in our survey area.  In addition, \dterm\ calibration would require determination of the primary beam Mueller matrices beyond our current accuracy. (Note that this is an ongoing effort; see Paper {\sc ii}, in prep.). The consequences of this limitation are discussed in the analysis of our power spectra in Section~\ref{sec:res}.  

As a intermediate measure compatible with these limitations, we therefore adopted a minimization of the phase difference between the $V_{ij}^{xy}$ and $V_{ij}^{yx}$ visibilities, minimizing a sum of squared weighted residuals $w$:

\begin{equation}
w(\nu,\,t,\,\tau_{xy}) = \sum_{ij} | V_{ij}^{xy} - V_{ij}^{yx}\exp(-2\pi i \nu \tau_{xy}) |^2
\end{equation}
to find an estimated value of $\tau_{xy}$ for the array at each $(\nu,\,t)$ sample. This is equivalent to assuming that the sky is intrinsically not circularly polarized at the frequencies observed by PAPER.

\begin{figure}
\includegraphics[scale=0.45]{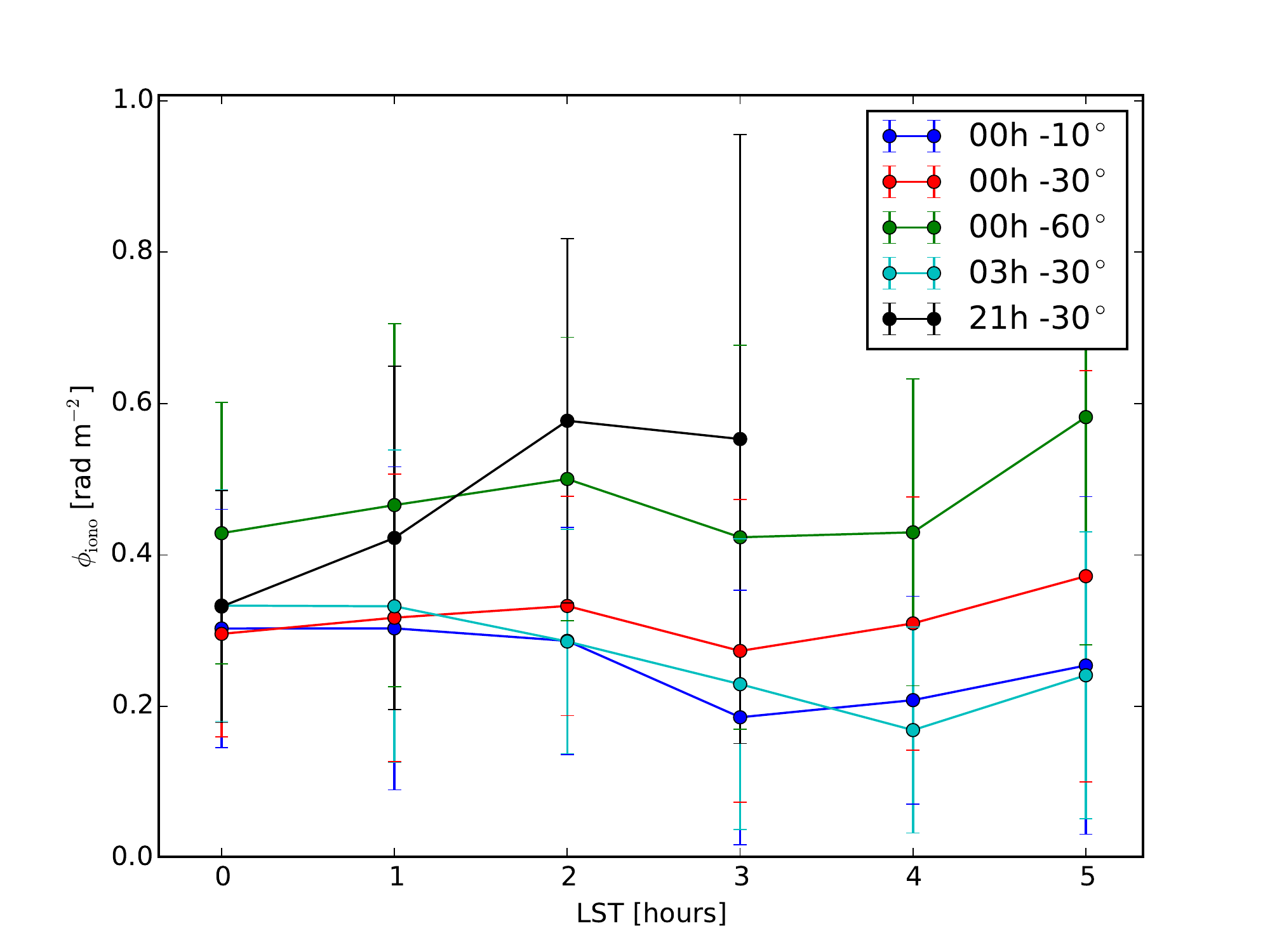}
\caption{The values of ionospheric RM for different lines of sight the range of LSTs in this analysis, as calculated by {\sc ionFR} \protect\citep{Sotomayor-Beltran.13}. The 21h,0$^{\circ}$ line of sight goes beneath the horizon after LST=3h, and therefore has fewer data points.\\ }
\label{fig:ionosphere}
\end{figure}

We choose not to correct for ionospheric Faraday rotation in our calibration. Not only is this difficult to do for widefield instruments, but also the ionosphere was relatively stable during the observations, so we expect little incoherent averaging during the power spectrum stage below. We calculated the stability of ionospheric RM ($\phi_{\rm iono}$) using the {\sc ionFR} software \citep{Sotomayor-Beltran.13}, which calculates the $\phi_{\rm iono}$ for a given longitude, latitude and time by interpolating values of GPS-derived total electron content maps and the International Geomagnetic Reference Field \citep{Finlay.10}. The values of $\phi_{\rm iono}$ for different lines of sight are shown in Figure~\ref{fig:ionosphere}. 
Fluctuations of $\phi_{\rm iono}$ will cause incoherent time-averaging and subsequent loss of polarized signal. Using the formalism of \citet{Moore.15} to calculate the attenuation factor, we found that none of the lines of sight (except for the 21h,0$^{\circ}$ one which goes beneath the horizon) shown are responsible for attenuating signal by $>\,20\%$ in power-spectrum space (see Section~\ref{subsec:create_pspec}).

We form linear combinations of the instrument visibilities, the so-called Stokes visibilities \citep[see e.g.][]{Moore.13} $V^I,\,V^Q,\,V^U$ and $V^V$ as:

\begin{equation}
\left(\begin{array}{c}
V^{I}\\
V^{Q}\\
V^{U}\\
V^{V}\end{array} \right)
=
\left( \begin{array}{cccc}
1 & 0 & 0 & 1 \\
1 & 0 & 0 & -1 \\
0 & 1 & 1 & 0 \\
0 & -i & i & 0 \end{array} \right) 
\left(\begin{array}{c}
V^{xx}\\
V^{xy}\\
V^{yx}\\
V^{yy}\end{array} \right) 
\label{eq:Stokes}
\end{equation}
Data that were reduced, calibrated, and formed into Stokes visibilities were separated into delay spectra inside and outside of the horizon for each baseline. We used a 50\,ns margin for what was considered `inside' the horizon, in order to confine all supra-horizon emission \citep[e.g.][]{Parsons.12a, Pober.13} to the foreground component of the data. We implemented a one-dimensional CLEAN \citep{Parsons.09, Parsons.12b} with a Blackman-Harris window to a tolerance of $10^{-9}$. RFI is more easily identified in foreground-removed data, so we RFI-flagged again on the background data deviations greater than $3\sigma$. We then added the inside- and outside-horizon visibilities back together; RFI flags were preserved in the process. 

\begin{figure}
\includegraphics[width=\columnwidth]{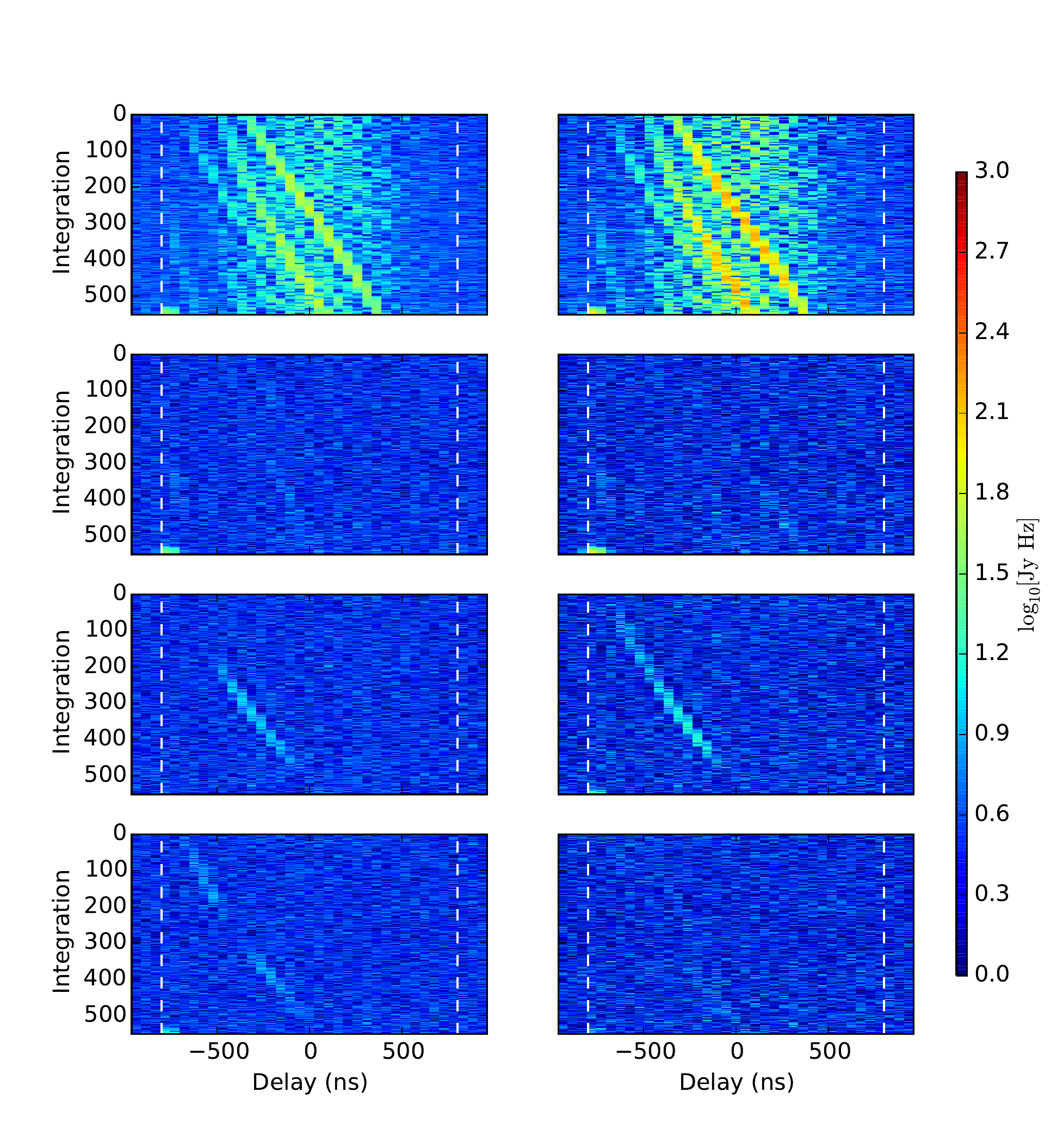}
\caption{The absolute value of delay-transformed visibilities over the bandwidth (146--166\,MHz) used to create the power spectra shown in this work. The left and right columns show the visibilities before and after absolute calibration (and for Stokes U and V, the application of the $\tau_{xy}$ parameter), respectively, for baseline formed by antennae 6 and 14 ($\sim 250$\,m in length, approximately East-West). The flux scale in the left column as been boosted for a more fair comparison to the absolute-calibrated data. From top to bottom, the rows correspond to Stokes I, Q, U and V. 
The horizon limit is marked by white dashed lines. 
}
\label{fig:delay_spectra}
\end{figure}

The effect of our calibration is shown in the delay-transformed visibilities in Figure~\ref{fig:delay_spectra}. As is apparent in Figure~\ref{fig:sky_image}, after improved calibration there are fewer delay tracks (i.e. sources) in the Stokes Q visibilities, while there is little overall change in Stokes U. The minimization of Stokes V, performed after the imaging calibration stage, moves power from Stokes V into Stokes U, effectively accounting for part of a \dterm\ correction. But without an accurate \dterm\ calibrator, Stokes U exhibits additional (and dominant) \dterm\ leakage from Stokes I, 
in this case due to Pictor A. Pictor A is the brightest source in Stokes I in our observed field, and thus dominates the visibility shown.  There is no reason to suppose that Pictor A is pure Stokes U (compare also Figure \ref{fig:sky_image}), and thus the bulk of this emission must be leakage.

\subsection{Creating power spectra}
\label{subsec:create_pspec}

Expressing the visibility $V_{ij}^{pq}(\nu,t)$ observed at time $t$ (see Equation~\ref{eq:visibility}) in terms of the geometrical delay $\tau_g=\vec{b}(t) \cdot \hat{s}(l,m)/c$ for the baseline $ij$, \citet{Parsons.12a} define the delay transform as the Fourier transform of the visibility along the frequency axis:
\begin{equation}
\tilde{V}_{ij}^{pq}(\tau,t) = \int d\nu \, V_{ij}^{pq}(\nu,t) e^{2\pi i \nu \tau}
\end{equation}

We can represent the power at each frequency and baseline in an array as a power spectrum in terms of their respective Fourier components $k_{\parallel}$ and $k_{\perp}$ as:
\begin{equation}
P(k_{\parallel},k_{\perp}) \approx |\tilde{V}_{ij}^{pq}(\tau,t)|^2 \frac{X^2Y}{\Omega B}\left(\frac{c^2}{2k_B\nu^2}\right)^2
\label{eq:pspec}
\end{equation}
where $B$ is the bandwidth, $\Omega$ is the angular area (i.e. proportional to the beam area), and X and Y are redshift-dependent scalars calculated in \citet{Parsons.12b}. 

To form $|\tilde{V}_{ij}^{pq}(\tau,t)|^2$, consecutive integrations were cross-multiplied, phasing the zenith of latter to the former i.e.:
\begin{equation}
|\tilde{V}_{ij}^{pq}(\tau,t)|^2 \approx |{V}_{ij}^{pq}(\tau,t)\times {V}_{ij}^{pq}(\tau,t+\Delta t)e^{i\theta_{ij,{\rm zen}}(\Delta t)}|^2
\end{equation}
where $\Delta t$=10.7 seconds and $\theta_{ij,{\rm zen}}(\Delta t)$ is the appropriate zenith rephasing factor. This method should avoid noise-biased power spectra except on very long baselines, which the PAPER configuration does not contain, while sampling essentially identical $k$-modes.  Note that this is the same method used by \citet{Pober.13} in their investigation of the unpolarized wedge.

\section{Results}
\label{sec:res}
Combining visibilities using Equation~\ref{eq:Stokes}, we formed power spectra over frequencies 146--166\,MHz according to Equation~\ref{eq:pspec} using consecutive integrations for each Stokes visibility over time, and gridded our results into $k$-space, averaging in time. 
The $k_{\perp}$-axis is binned with a resolution of $4.65\times10^{-4}\,h$\,Mpc$^{-1}$ to slightly reduce gaps in $k$-space due to missing baselines. This gave an average bin occupancy of 1.7$\pm$0.9.
The resolution in $k_{\parallel}$ ($5.06\times10^{-4}\,h$\,Mpc$^{-1}$) is set by the 20\,MHz bandwidth with 500\,kHz resolution that we use in this analysis.
Note that the Blackman-Harris window used in the delay-filtering stage after forming Stokes visibilities correlates adjacent frequency bins, and hence $k_{\parallel}$ bins.
Each $(k_{\perp},k_{\parallel})$ bin was normalized by its occupancy.

Two-dimensional power spectra have been proven as powerful tools for large dipole array experiments, not only for assessing cosmology but also in order to constrain instrumental and analytical systematics \citep[e.g.][]{Morales.12}. Polarization axes are a useful addition for such analyses, since we expect Stokes I to be approximately 3 orders of magnitude stronger than the other polarization products at the low radio frequencies and tens-of-arcminute scales native to PAPER observations \citep[e.g.][]{Pen.09, Moore.13} and when observing far from the Galactic Plane. This alone allows us to assume that much of the structure in the power spectra with power comparable to Stokes I is leakage. As we explore below, these leakage terms can come from direction-dependent effects \citep[e.g. wide-field beam leakage;][]{Carozzi.09} or direction independent ones \citep[e.g. Mueller matrix mixing via gain errors and D-terms;][]{TMS} and appear with high signal-to-noise in power spectra.

\begin{figure*}
\centering
\includegraphics[scale=0.45]{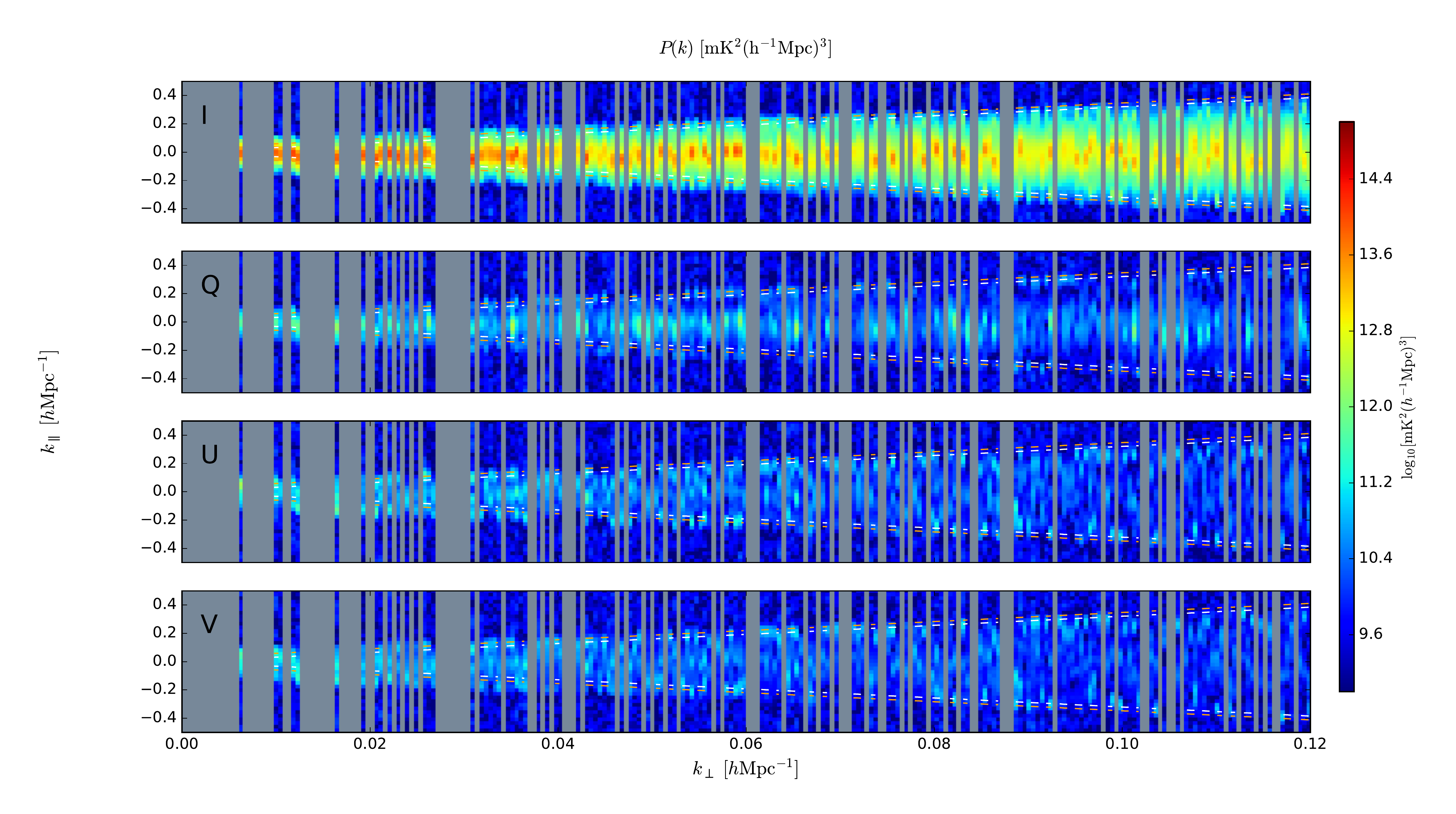}
\noindent\rule{14cm}{0.6pt}
\includegraphics[scale=0.45]{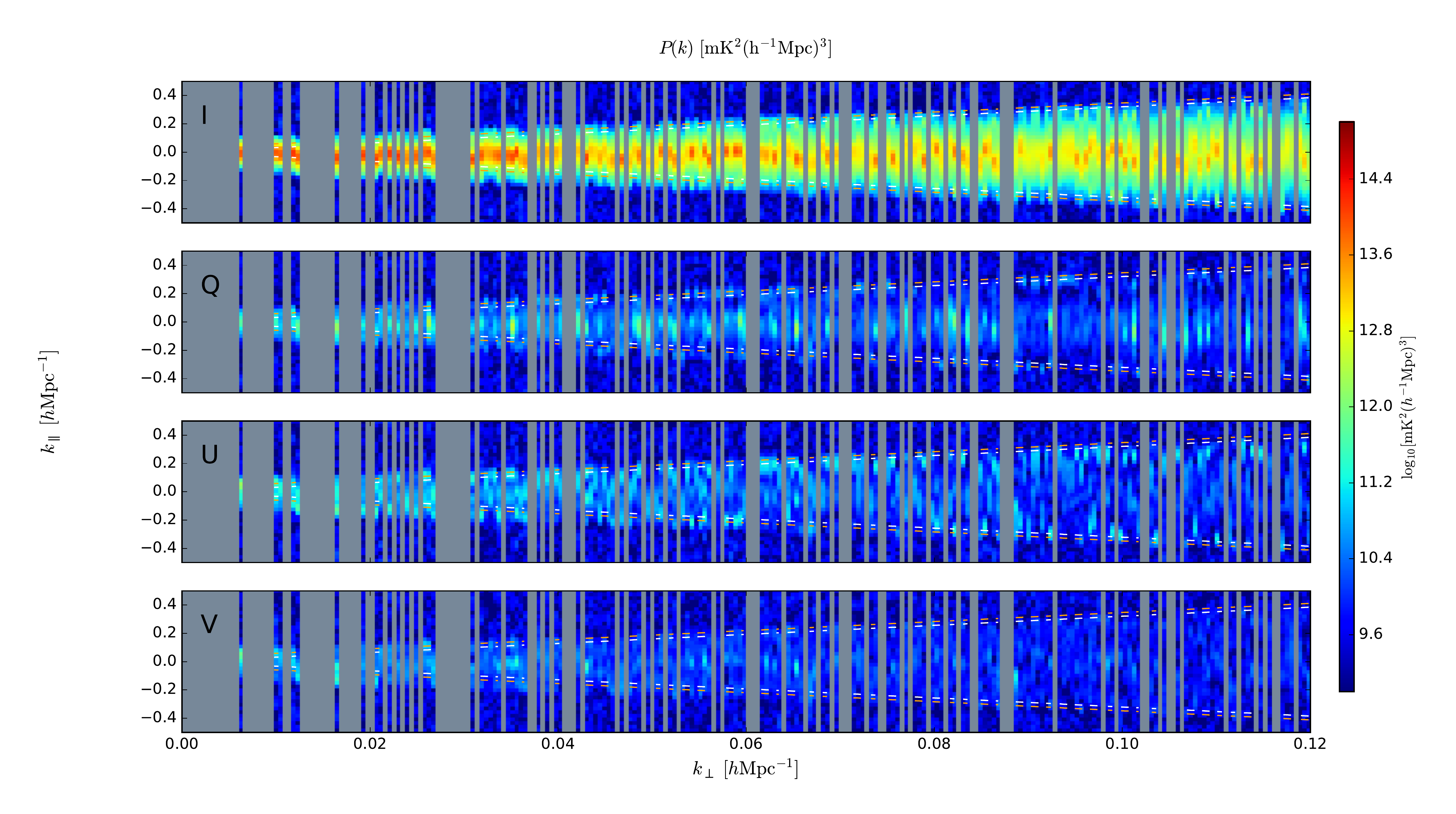}
\caption{\textit{Above:} Log-scaled 2D power spectra formed from (\textit{top to bottom}): I, Q, U and V visibilities after before absolute calibration. Blank regions indicate the incomplete $uv$ coverage for a given $k_{\perp}(u)$. 
The colorbar spans $10^{9}$ to $10^{15}$ mK$^2$($h^{-1}$Mpc)$^3$. The flux scale has been boosted for a more fair comparison to the absolute-calibrated data.
\textit{Below:} The same organization as above, but after absolute calibration.
Briefly, the structure in Stokes I is consistent with a point-source-dominated field with a weak diffuse component. The other Stokes parameters are consistent with calibration errors and systematics: Stokes Q shows gain errors on specific antennae, Stokes U gives an estimate of possible D-term leakage, and any structure in V shows unaccounted-for systematics, due to D-terms or mis-oriented antennae.}
\label{fig:forks}
\end{figure*}

\begin{figure*}
\includegraphics[scale=0.45]{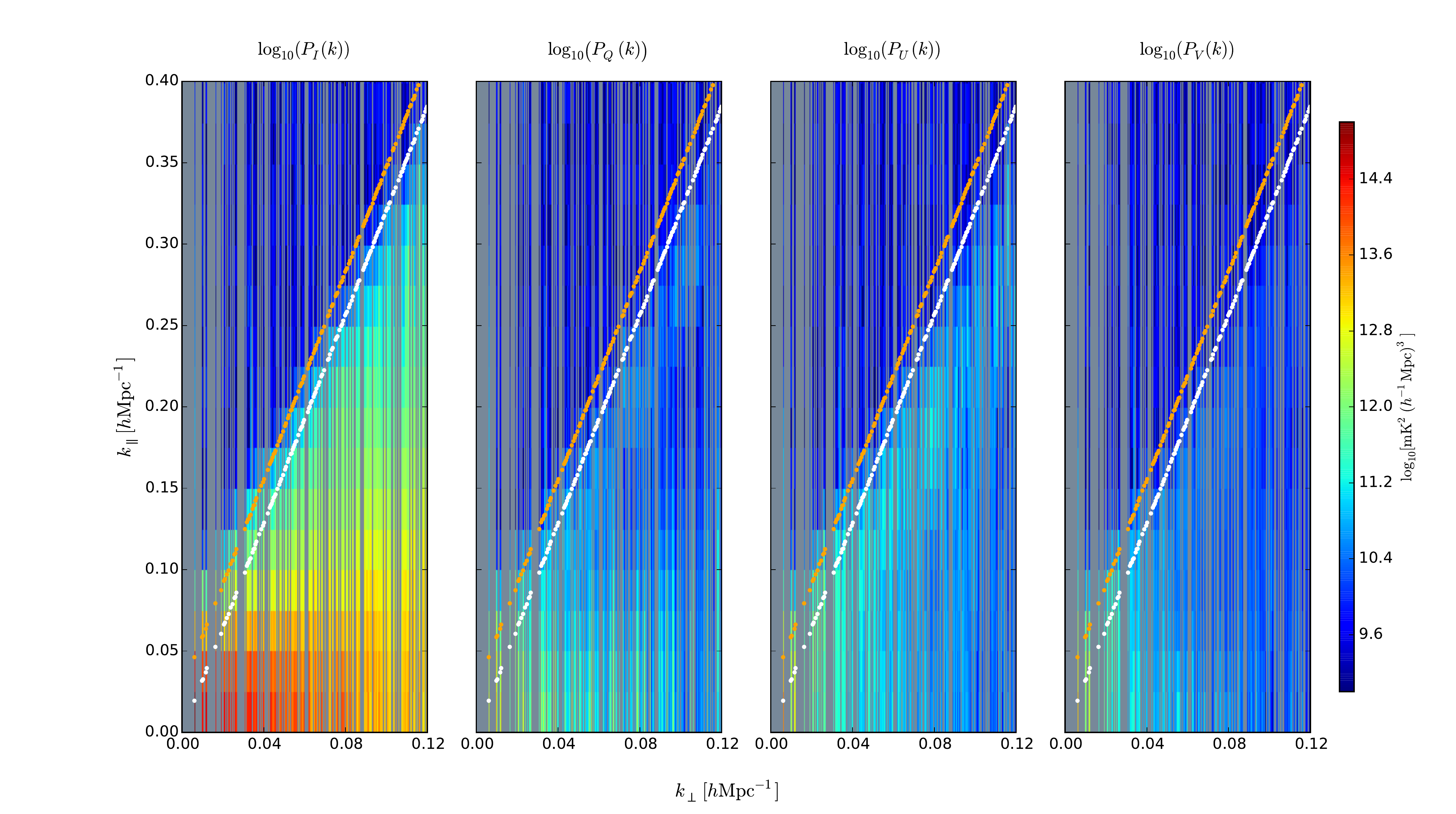}
\caption{Log-scaled 2D power spectra formed from (\textit{left to right}): I, Q, U and V absolute-calibrated visibilities. Blank regions indicate the incomplete $uv$ coverage for a given $k_{\perp}(u)$. White and orange lines indicate the horizon and horizon plus a 50\,ns boundary for super-horizon emission. }
\label{fig:wedges}
\end{figure*}

Figure~\ref{fig:forks} shows power spectra in `pitchfork' form \citep{Nithya.15b,Nithya.15a}, with $k_{\parallel}$ in negative and positive directions (according to the East and West horizons, marked in white (horizon) and orange (horizon+50\,ns delay, respectively). Each Stokes parameter pitchfork has its own interesting characteristics, which allow us to analyze different sky and instrument behaviors. The `wedges' described in the literature that define the EoR window are simply the average of negative and positive values of $k_{\parallel}$. While we focus on the pitchfork expression of the power spectra in our results, we also show them in wedge form in Figure~\ref{fig:wedges}\footnote{Note the difference in the power distribution within the horizon differs from that shown in the \citet{Pober.13} $V^{yy}$ wedge. That study used the PAPER 64-element, single-polarization imaging array to create power spectra in a `loud' field containing point sources and Galactic signal, causing their wedge to be `fuller' than the ones presented in this study.}.

Simplifying the results of \citet[][see their papers for a full discussion]{Nithya.15b,Nithya.15a}, we expect power from diffuse emission to appear at low values of $k_{\perp}$ and high values of $k_{\parallel}$, while point sources lie at all $k_{\parallel}$ (all over the sky) but are down-weighted by the primary beam, which is broad, leaving a concentration of the power close to the $k_{\parallel}=0$ line. 

In Stokes I, we see the strongest power on most baselines arising at values $k_{\parallel}\approx 0$. This is expected in a situation of point sources that are relatively bright compared to any diffuse emission. Indeed, at the LSTs we observed at, several unresolved bright point sources transit the field (e.g. Figure~{\ref{fig:sky_image}}), while the dominant source of diffuse emission at these frequencies, the Galactic plane, was below the horizon. However, we do see strong super-horizon emission at $0.02 \leqslant k_{\perp} \leqslant 0.03$, biased towards negative $k_{\parallel}$ values. There is also a decrease in power with increasing $k_{\perp}$ -- both of these effects are consistent with the \citet{Nithya.15a} simulations of faint diffuse structure transiting zenith. 

The Stokes Q wedge shows a concentration of power close to $k_{\parallel}\approx 0$, similar to Stokes I.
The inherent low polarization fraction at our frequencies works in our favor in detecting gain errors, since Stokes Q is largely expected to be faint, and thus the gain errors causing leakage from I appear at high signal-to-noise there.
Indeed, this power decreases noticeably with more accurate gain amplitude calibration, but bright streaks at specific values of $k_{\perp}$ remain, suggesting lower-level residual gain calibration errors on select baselines.  Another possible source of power in Stokes Q stems from widefield direction-dependent gain errors causing a non-smooth evolution of the sources on the edges of the beam. However, we would expect this effect to be biased towards horizon values of $k_{\parallel}$.

Power appears distributed in `pockets' in the Stokes U power spectrum, not strongly correlated with the distribution of power in I. Stokes I is able to leak into Stokes U via \dterm\ leakage \citep{TMS, Geil.11}, which could occur at any post-amplification stage of observations, such as in cables or receivers. These leakages would be direction independent, and therefore uncorrelated in $k$-space. Such a mechanism could explain the behavior within Stokes U wedge. Before absolute calibration, similar structure is seen in the Stokes V power spectrum.

At these frequencies, Stokes V is thought to be intrinsically zero, with few exceptions. However, \citet{HBS-1} show that antennae rotated with respect to one another can produce erroneous Stokes V power via I$\rightarrow$V leakage.\footnote{
It should be noted that while such an error could plausibly have been made in the antenna placement for this imaging array, it is extremely unlikely that it would be made in the redundant PAPER configuration for EoR seasons. In these cases, the antennae were positioned to sub-cm accuracy.} This effect may explain some of the small pockets of power that remain in the Stokes V power spectrum after absolute calibration, although such an effect is also consistent with \dterm\ leakage. The fact that power within the horizon is greater than the noise level may also be due to I $\to$ V leakage through the primary beam.

\begin{figure*}
\centering
\includegraphics[scale=0.45]{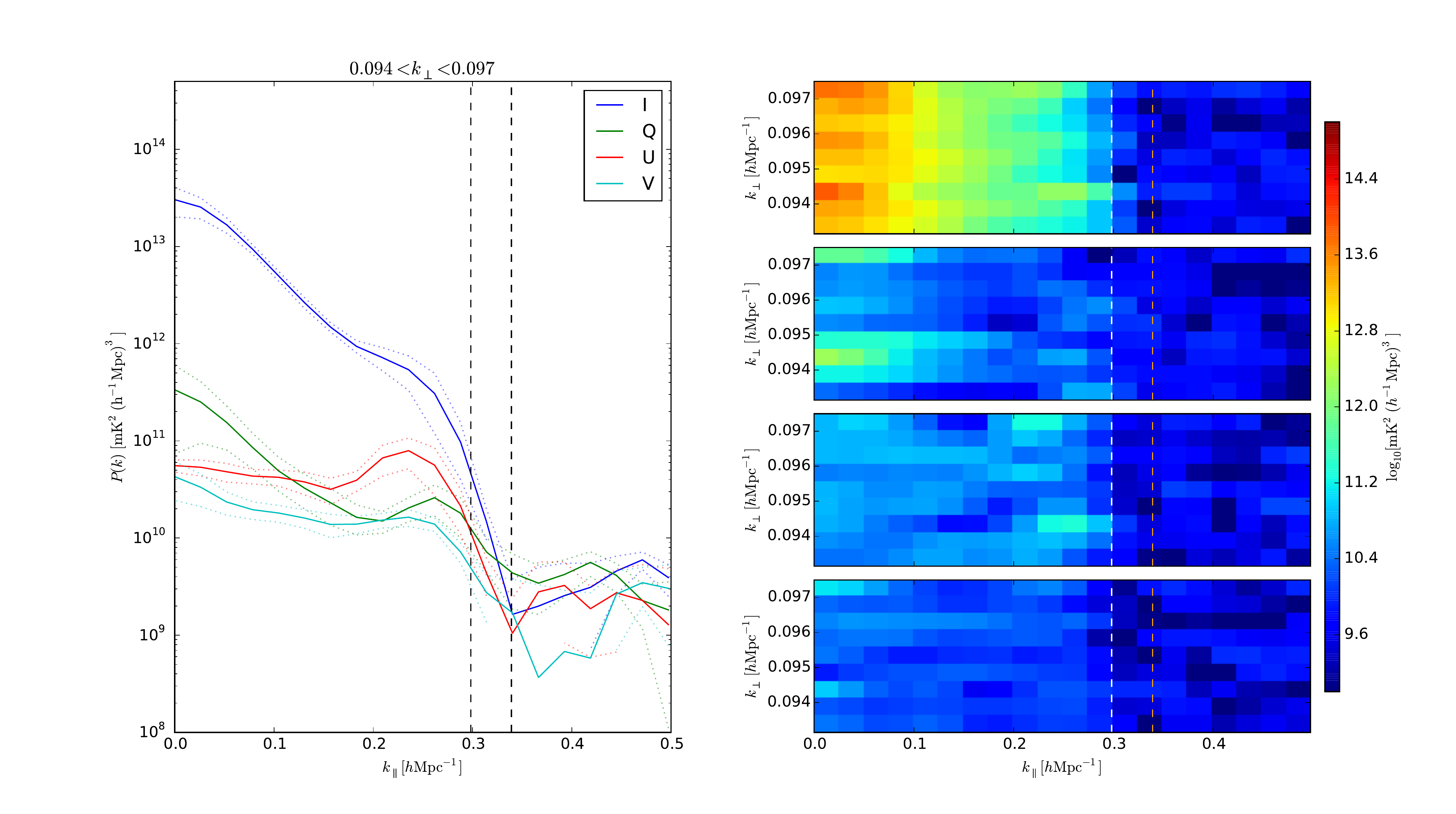}\\
\caption{\textit{Left}: The average power in $0.093\,<\,k_{\perp}\,<\,0.098\,h{\rm Mpc^{-1}}$ as a function of $k_{\parallel}$ for each polarization.  In the horizon--super-horizon range that we see the sharp fall-off in power indicative of the edge of the EoR window. 
Outside of the wedge, Stokes I, Q and U are at noise level, while Stokes V is below noise level, most likely due to the $\tau_{xy}$ calibration scheme removing a degree of freedom from this Stokes parameter. \textit{Right}: the region of k-space that was averaged over to create the lines in the left panel. From top to bottom, the panels correspond to Stokes I, Q U and V.}
\label{fig:high_kperp_slice}
\end{figure*}

The relationship between polarizations is highlighted in Figure~\ref{fig:high_kperp_slice}.
We show a slice of the wedges over $0.097<\,k_{\perp}\,<0.098\,h{\rm Mpc^{-1}}$ ($\sim$175\,m) for Stokes I, Q, U and V (right panels) and the average power over these slices as a function of $k_{\parallel}$ (left panel). The standard deviations for each Stokes parameter are shown as dotted lines. 
Dashed vertical lines show the horizon at $k_{\perp}$=0.097 (left) and super-horizon at $k_{\perp}$=0.098 (right).

A heartening aspect of Figure~\ref{fig:high_kperp_slice}, and indeed all of the power spectra in this work, is that the power in Stokes Q, U and V proves to be just as confined within the horizon as Stokes I. Whether the polarized Stokes parameters are due to real polarization or miscalibration, not enough spectral structure is being introduced to move emission into the EoR window. Outside of the horizon, Stokes I, Q and U are consistent with the noise level expected for this range of $k$-modes ($P_{\rm noise}\sim10^{9}$ mK$^2$($h^{-1}$Mpc)$^3$), according to the formalism \citet{Parsons.12b} and assuming a system temperature $T_{\rm sys}=450\,$K \citep[e.g.][]{Moore.15}.

\section{Discussion and Conclusions}
\label{sec:disc}

We have presented measurements of instrumental polarization leakage in PAPER-32 using 2D power spectra. These have allowed us to quantify some of the possible instrumental effects that could limit a statistical detection of the EoR within the wedge, diagnosed in the Fourier space most relevant EoR statistical detection experiments. To our knowledge, this is the first study of Q, U and V 2D power spectra at these $k$-modes. We have shown that power from Stokes Q, U and V is as confined to the wedge as Stokes I. Any calibration errors do not appear to spread power outside the horizon. 

In their study of 2D power spectra, \citet{Asad.15} reported evidence of polarized leakage into the EoR window at the sub-percent level, considering a $4^\circ$ degree field of view. Their study differs from this work not only over the field of view ($4^\circ$ versus almost whole-sky), but also in the observing mode (tracked versus drift scan) and in the different k-space probes by LOFAR's longer baselines. In this work the power spectrum is calculated on a per-baseline basis, whereas their study calculates power spectra for gridded data which are more prone to mode mixing effects \citep{Hazelton.13}.

Our results are expected, in principle, to be more prone to leakage contamination due to the intrinsic extremely wide field of view of the PAPER primary beams, however, we see no evidence of leakage in the EoR window down to our sensitivity limits even without correction for polarized beams that is instead included in \cite{Asad.15}. Our analysis indicates therefore that neither intrinsic polarized emission nor the PAPER primary beam are leaking power in the EoR window, although longer integrations are required to demonstrate that this is true down to the sensitivities required for EoR detection. 

We showed that systematics can be probed with high signal-to-noise using 2D polarized power spectra, using the inherently low polarization fraction at the frequencies PAPER observes at to our advantage. We found that gain errors on specific baselines were easily probed using Stokes Q power spectra. Gain errors appear as continuous streaks within the horizon at specific values of $k_{\perp}$, allowing us to diagnose the precision of the gain calibration on a per-baseline basis. This is much more difficult to do with only Stokes I power spectra in a non-redundant array, and can be accomplished quickly without imaging. While the features in the Stokes U power spectra are more difficult to attribute to specific baselines, they appear to be consistent with direction-independent leakage. 
Stokes V power is slightly higher than noise-level within the horizon, suggesting a small but unaccounted-for leakage term from Stokes I, an effect which will be explored in Paper~{\sc ii}.

The measurements presented here came from PAPER 32-element data taken in 2011. PAPER has since placed limits on the EoR power spectrum using redundant-array 32-element \citep[e.g.][]{Parsons.14} and 64-element data \citep[e.g.][]{Ali.15}. \citet{Moore.15} showed limits on the polarized power spectrum with 32-element data. Limits on the polarized power spectrum from PAPER-64 and PAPER-128 are currently being worked on, which will allow us to more fully quantify the effect of polarized foregrounds on a statistical detection of the EoR. 
The PAPER-64 array was a purely redundant grid, but PAPER-128 contains outrigger antennae that provide greater $uv$ coverage than the data we present in this work. A future paper will show power spectra from this array, which can leverage {\sc omnical}'s precise redundant calibration for most of the antennae and more than 1000 hours of integration time to produce extremely deep measurements of polarization. 

\section*{Acknowledgments}
The authors would like to thank Adam Beardsley, Bryna Hazelton, Zachary Martinot, Miguel Morales, and Nithyanandan Thyagarajan for helpful conversations, and the anonymous referee for their helpful comments.

We thank SKA-SA for their efforts in ensuring the smooth running of PAPER. PAPER is supported through the NSF-AST program (awards 0804508, 1129258, and 1125558), the Mt. Cuba Astronomical Association, and by significant efforts by staff at NRAO.
This work is based on research supported by the National Research Foundation under grant 92725. Any opinion, finding and conclusion or recommendation expressed in this material is that of the author(s) and the NRF does not accept any liability in this regard. The financial assistance of the South African SKA Project (SKA SA) towards this research is hereby acknowledged. Opinions expressed and conclusions arrived at are those of the authors and are not necessarily to be attributed to the SKA SA. CDN is supported by the SKA SA scholarship program. 
ARP acknowledges support from the University of California Office of the President Multicampus Research Programs and Initiatives through award MR-15-328388, as well as from NSF CAREER award No. 1352519, NSF AST grant No.1129258, and NSF AST grant No. 1440343.

\bibliography{wedgebib}{}

\end{document}